\documentclass[10pt]{article}
\usepackage[pctex32]{graphics}
\usepackage[dvips]{graphicx}

\usepackage{epsfig}
\usepackage{graphicx}
\usepackage{indentfirst}
\usepackage{amsmath}
\usepackage{amsfonts}
\usepackage{amssymb}
\usepackage{hyperref}
 \usepackage{latexsym}
\usepackage{color}

\begin{document}
\begin{center}
{\Large\bf   $\Lambda$CDM Model in $f(T)$ Gravity: Reconstruction, Thermodynamics and Stability }\\

\medskip

I. G. Salako$^{(a)}$\footnote{e-mail:ines.salako@imsp-uac.org}, M. E. Rodrigues$^{(b,c)}$\footnote{e-mail: esialg@gmail.com}, A. V. Kpadonou$^{(a,d)}$\footnote{e-mail: vkpadonou@gmail.com}, 
M. J. S. Houndjo$^{(a,e)}$\footnote{e-mail:
sthoundjo@yahoo.fr} and J. Tossa$^{(a)}$\footnote{e-mail: joel.tossa@imsp-uac.org}

$^a$ \,{\it Institut de Math\'{e}matiques et de Sciences Physiques (IMSP)}\\
 {\it 01 BP 613,  Porto-Novo, B\'{e}nin}\\
$^{b}$\,{\it\ Faculdade de F\'{\i}sica, Universidade Federal do Par\'{a}, 66075-110, Bel\'em, Par\'{a}, Brazil}\\
$^{c}$\,{\it     Faculdade de Ci\^encias Exatas e Tecnologia, Universidade Federal do Par\'a - Campus Universit\'ario de\\
Abaetetuba, CEP 68440-000, Abaetetuba, Par\'a, Brazil.      }\\
$^{c}$\, {\it Ecole Normale Sup\'erieure de Natitingou - Universit\'e de Parakou - B\'enin}\\ 

$^{d}$\,{\it Facult\'e des Sciences et Techniques de Natitingou - Universit\'e de Parakou - B\'enin} \\

\date{}

\end{center}
\begin{abstract}
We investigate some cosmological features of the $\Lambda CDM$ model in the framework of the generalized teleparallel theory of gravity $f(T)$ where $T$ denotes the torsion scalar. Its reconstruction is performed giving rise to an integration constant $Q$ and other input parameters according to which we point out more analysis. Thereby, we show that for some values of this constant, the first and second laws of thermodynamics can be realized in the  equilibrium  description,  for the universe with the temperature inside the horizon equal to that at the apparent horizon. Moreover, still within these suitable values of the constant, we show that the model may be stable using the de Sitter and Power-Law cosmological solutions.
\end{abstract}

Pacs numbers: 98.80.-k, 04.50.Kd, 05.70.Ln

\section{Introduction}
It is strongly known nowadays  that our universe is experiencing an accelerated expansion supported by many cosmological observational data such as type Ia supernovae \cite{debamba1}, cosmic microwave background (CMB) radiation \cite{debamba2}-\cite{debamba3}, large scale structure \cite{debamba4}, baryon acoustic oscillations \cite{debamba5}, and weak lensing \cite{debamba6}.  A possible responsible for this late-time acceleration is the so-called dark energy with negative pressure. An alternative approach for understanding this strange component of the universe is modifying the  standard theories of gravity, namely, General Relativity (GR) or Teleparallel Theory Equivalent to GR (TEGR)\cite{a'}. Among several modified theories of gravity ( $f(R)$, $f(R,\mathcal{T})$ \cite{ma1}-\cite{ma6}, $f(G)$) \cite{mj1}-\cite{mj5}, where $R$ is the curvature scalar, $\mathcal{T}$ the trace of the energy momentum tensor, $G$ the invariant of Gauss-Bonnet defined as
 $G=R^2-4R_{\mu\nu}R^{\mu\nu}+ R_{\mu\nu \lambda\sigma}R^{\mu\nu\lambda\sigma}$, 
 special attention is attached to the so-called $f(T)$ gravity as the modified version of the TEGR, where $T$ denotes the torsion scalar. Several works have been developed in the framework of this modified theory of gravity and interesting results have been found \cite{st1}- \cite{st39}.\par
The fundamental  connection between gravitation and thermodynamics comes from the study of black hole thermodynamics \cite{bambaa1}-\cite{bamba'}. In the framework of GR,  Clausius relation in thermodynamics yields the Einstein equation with the proportionality of the entropy to the horizon area \cite{bamb1}. The technique used to explain this relation is also extended to other gravitational theories, namely, modified theories of gravity and the thermodynamics laws are view as generalized thermodynamics laws (because of their analytical deviation from the GR  \cite{manu1,manu2}). The well known modified theory of gravity $f(R)$ has received more attention on this way. Thereby, it has been shown that the gravitational field equation from the Clausius approach is obtained through the non-equilibrium aspect of thermodynamics \cite{bamb2}-\cite{bamb3}. \par
In the view of this feature of thermodynamics in $f(T)$ gravity, we attach our attention to the so-called $\Lambda CDM$ model. Note that $\Lambda$CDM is an interesting model because of its particularity in explaining the present stage of the universe. This model has been studied in several works in the frameworks of other types of modified theory of gravity, but for other purposes. \par 
  The potential works that have relationships with our manuscript, that is, dealing with thermodynamics, stability and $\Lambda$CDM model in $f(T)$, are that of Bamba \cite{bamba'} and setare \cite{st6}. Indeed, Bamba and collaborators widely explored thermodynamics of the apparent horizon in  f(T) gravity with both equilibrium and non-equilibrium descriptions \cite{bamba'}, obtaining interesting results. Still in the framework of $f(T)$ gravity, Bamba and collaborators, in their interesting paper entitled `` Reconstruction of $f(T)$ gravity: Rip cosmology, finite-time future singularities and thermodynamics" \cite{st15},  in the section $IV$ and subsection ${\bf C2}$, described $\Lambda$CDM model by consideration the algebraic function action function of the form $f(T)=T-2\Lambda$, with $\Lambda>0$, which is a possibility for describing $\Lambda$CDM model as is usually done in General Relativity (GR). Thermodynamics is also developed in that paper, but about the finite future singularities models. Setare and collaborator \cite{st6} performed the reconstruction of the algebraic function $f(T)$ assuming that the background cosmic history is provided by flat $\Lambda$CDM, also assumed having the same form as in Bamba's paper.\par
Nowhere, these papers undertook at the same time the study of thermodynamics and stability of $\Lambda$CDM model. This is the task of our paper. More precisely, in this paper, we reconstruct the algebraic function according to the $\Lambda$CDM feature in a generic scheme. This algebraic function presents a positive integration constant $Q$, suitably determined according an initial condition.  It is important to point out that one can clearly see from  our model, Eq. (24) of our manuscript,  that the action algebraic used in the Bamba's paper is recovered by setting $Q=0$ from the one of our manuscript. Therefore, the model presented in this paper is more general than the ones used by Bamba and Setare,  in the description of $\Lambda$CDM feature.    
\par
After reconstructing the algebraic  function $f(T)$ according to $\Lambda CDM$ properties \cite{lambda}, we check the occurrence of the first and second laws of thermodynamics according to the input parameters.   We undertake in this work the equilibrium description of the thermodynamics, where we assume that the temperature of the universe inside the horizon is equal to that at the apparent horizon. We emphasize that the possibility of the second law of the thermodynamics being realized in the equilibrium description in the framework of $f(T)$ gravity has been shown by Bamba and collaborator \cite{bamba'}, and followed in this work.
We find that the  first and second laws of thermodynamics may be satisfied and the comments are presented in a upcoming section.  \par
Another important feature which has been investigated
in this work is the stability of the model under consideration. By 
this way, we consider two interesting cosmological solutions (de Sitter
and power-law solutions) and analyse the constraints on the input parameters 
for obtaining the stability  of this $\Lambda CDM$ model 
and we find that stability is always realized within the de Sitter solutions, while for the power-law solutions stability is obtained for $0< Q<  2.94\times 10^{-42} GeV$ 
and $\alpha >1/3(1+w_m)$. \par
The paper is organized as follows. In Sec. \ref{sec2}, we present
the generality of the theory within FLRW cosmology. The reconstruction of $\Lambda CDM$ is performed in Sec. \ref{sec3}. The Sec. \ref{sec4} is devoted to the study of the first and second laws of thermodynamics. The stability of the model is analysed in Sec. \ref{sec5} and the conclusion is presented in Sec. \ref{sec6}

\section{Generality on $f(T)$  gravity within  FLRW Cosmology}\label{sec2}

The modified version of the Teleparallel gravity is that for which the torsion in the action is substituted by an arbitrary function depending on the torsion scalar. As well as in teleparallel theory and its modified version, gravity  is described using orthonormal tetrads components which are defined in the tangent space at each point of the manifold. The line element can be written as 
\begin{eqnarray}
ds^2=g_{\mu\nu}dx^\mu dx^\nu=\eta_{ij}\theta^i\theta^j\,,
\end{eqnarray}
with the definition
\begin{eqnarray}
d^\mu=e_{i}^{\;\;\mu}\theta^{i}; \,\quad \theta^{i}=e^{i}_{\;\;\mu}dx^{\mu}.
\end{eqnarray}
Here, $\eta_{ij}=diag(1,-1,-1,-1)$ is the Minkowskian metric and $\{e^{i}_{\;\mu}\}$ the components  of the tetrad satisfying the following identity
\begin{eqnarray}
e^{\;\;\mu}_{i}e^{i}_{\;\;\nu}=\delta^{\mu}_{\nu},\quad e^{\;\;i}_{\mu}e^{\mu}_{\;\;j}=\delta^{i}_{j}.
\end{eqnarray}
Instead of the Levi-Civita's connection in the general relativity and its modified version, the teleparallel theory and its modified versions are governed by the Weizenbock's connection, defined by
\begin{eqnarray}
\Gamma^{\lambda}_{\mu\nu}=e^{\;\;\lambda}_{i}\partial_{\mu}e^{i}_{\;\;\nu}=-e^{i}_{\;\;\mu}\partial_\nu e_{i}^{\;\;\lambda}.
\end{eqnarray}

From this connection one can now determine the main geometrical objects. The first is the torsion, defined as
\begin{eqnarray}
T^{\lambda}_{\;\;\;\mu\nu}= \Gamma^{\lambda}_{\mu\nu}-\Gamma^{\lambda}_{\nu\mu},
\end{eqnarray}
from which we define the contorsion as
\begin{eqnarray}
K^{\mu\nu}_{\;\;\;\;\lambda}=-\frac{1}{2}\left(T^{\mu\nu}_{\;\;\;\lambda}-T^{\nu\mu}_{\;\;\;\;\lambda}+T^{\;\;\;\nu\mu}_{\lambda}\right)\,\,.
\end{eqnarray}
The above objects (torsion and contorsion) are used to define a new tensor $S_{\lambda}^{\;\;\mu\nu}$ as
\begin{eqnarray}
S_{\lambda}^{\;\;\mu\nu}=\frac{1}{2}\left(K^{\mu\nu}_{\;\;\;\;\lambda}+\delta^{\mu}_{\lambda}T^{\alpha\nu}_{\;\;\;\;\alpha}-\delta^{\nu}_{\lambda}T^{\alpha\mu}_{\;\;\;\;\alpha}\right)\,\,.
\end{eqnarray}

Using this later and the torsion one defined the torsion scalar as
\begin{eqnarray}
T=T^{\lambda}_{\;\;\;\mu\nu}S^{\;\;\;\mu\nu}_{\lambda}\,.
\end{eqnarray}

Since we are dealing with a modified version of the teleparallel gravity, a general algebraic function of the torsion is used, instead of just the torsion as in teleparallel gravity, and the action is written as

\begin{eqnarray}
 S= \int e \left[\frac{f(T)}{2\kappa^2} +\mathcal{L}_{m} \right]d^{4}x   \label{eq9}\,,
\end{eqnarray}
where $\kappa^{2} = 8 \pi G $ is the usual gravitational
coupling constant. By varying the action (\ref{eq9}) with respect
to the tetrads, one gets the following equations
of motion [\cite{a1}, \cite{a2}, \cite{a3}]
\begin{eqnarray}
S^{\;\;\; \nu \rho}_{\mu} \partial_{\rho} T f_{TT} + 
[e^{-1} e^{i}_{\;\; \mu}\partial_{\rho}(e e^{\;\; \mu}_{i}S^{\;\;\; \nu\lambda}_{\alpha} )
+T^{\alpha}_{\;\;\; \lambda \mu}   S^{\;\;\; \nu \lambda}_{\alpha} ]f_{T}+
\frac{1}{4}\delta^{\nu}_{\mu}f=\frac{\kappa^{2}}{2} \mathcal{T}^{\nu}_{\mu}  \label{eq10}\,,
\end{eqnarray}
where $\mathcal{T}^{\nu}_{\mu}$ is the energy 
momentum tensor, $f_{T} = df(T)/dT$ and 
$f_{TT}  = d^{2}f(T)/dT^{2}$ the first and second derivative of $f(T)$ with respect to $T$. These equations
clearly depend on the choice of the set of tetrads as shown in \cite{a4}.
Here we are interested in studying
flat $FLRW$ cosmologies, whose metric can be described by,
\begin{eqnarray}
 ds^{2}= dt^{2} - a^{2}(t)\left(dx^2+dy^2+dz^2\right)\,.  \label{eq10'}
 \end{eqnarray}
Let us now consider a set of diagonal tetrads related to the above metric as 
\begin{eqnarray}
 \{e^{a}_{\;\; \mu}\}= diag[1,a,a,a]. \label{eq11}
\end{eqnarray}
The determinant of the matrix (\ref{eq11}) is $e = a^{3}$. The components of the torsion tensor and contorsion
tensor are given by
\begin{eqnarray}
 T^{1}_{\;\;\; 01}= T^{2}_{\;\;\; 02}=T^{3}_{\;\;\; 03}=\frac{\dot{a}}{a},\\
 K^{01}_{\;\;\;\;1}=K^{ 02}_{\;\;\;\;2}=K^{ 03}_{\;\;\;\;3}= \frac{\dot{a}}{a},  \label{eq12}
\end{eqnarray}

and the components of the tensor $S^{\;\;\; \mu\nu}_{\alpha}$  are
\begin{eqnarray}
S^{\;\;\; 11}_{0}=S^{\;\;\; 22}_{0}=S^{\;\;\; 33}_{0}=\frac{\dot{a}}{a}.  
\end{eqnarray}
Therefore, the torsion scalar is calculated,  yielding 
\begin{eqnarray}
 T= -6H^{2}, \label{m1}
\end{eqnarray}
where $H=\dot{a}/a$ denotes the Hubble parameter. We assume now that the ordinary content of the universe is a perfect fluid with the equation of state $p_{m} = w_{m} \rho_{m}$, such that the energy-momentum tensor is given
by 
\begin{eqnarray}
\mathcal{T}^{\nu}_{\mu} =
diag(1,w_{m}, w_{m},
w_{m} ) \rho_m. 
\end{eqnarray}
By making use of the above quantities, the field equations
(modified Friedmann equations
 \ref{eq10}) read,
\begin{eqnarray}
 -Tf_T+\frac{1}{2}f&=& \kappa^{2} \rho_{m} \label{ines18}\,, \\
 2\dot{T}Hf_{TT }+2\left(\dot{H}+3H^2\right)f_T+\frac{1}{2}f&=&-\kappa^2 w_m\rho_m \label{ines18'}\,.
\end{eqnarray}
Making use of these equations, one easily get the  equation of continuity
\begin{eqnarray}
 \dot{\rho}_{m} + 3H(1+w_{m})\rho_{m}=0. \label{b1}
\end{eqnarray}
This equation is easily solved yielding the following solution
\begin{eqnarray}
\rho_{m}=\rho_0a^{-3(1+w_m)},
\end{eqnarray}
where $\rho_0$ denotes the current value of the energy density when the scale factor is set to unit.

\section{Reconstructing of $\Lambda CDM$ model in the framework of  $f(T)$ gravity}\label{sec3}

In this section, we are interested to the $f(T)$ model able to reproduce $\Lambda CDM$ feature. To do so, let us 
write the first generalized Friedmann equation in the following form
\begin{eqnarray}
 3H^{2}= \kappa^{2}\rho_{m} + \Lambda \,.\label{ines22}
\end{eqnarray}
 The above equation can be rewritten as,
\begin{eqnarray}
\frac{H^2}{H_0^2}=\Omega_m^0\ a^{-3(1+w_m)}+\Omega_{\Lambda}^0\ .
\label{lcdm3}
\end{eqnarray}
Here we have defined the usual cosmological parameters
\begin{eqnarray}
 \Omega_m^0=\frac{\rho_0}{\frac{2}{\kappa^2}H_0^2},\quad \Omega_{\Lambda}^0=\frac{\Lambda}{3H_0^2},
\end{eqnarray} 
  with $H_0$ being the value of the Hubble parameter today. Then, by using the expression of the torsion scalar (\ref{m1}), the scale factor  can be rewritten in terms of $T$, as
\begin{eqnarray}
a^{-3(1+w_m)}=-\frac{1}{6\Omega_m^0H_0^2}\left(T+6\Omega_{\Lambda}^0H_0^2\right)\ ,
\label{lcdm4}
\end{eqnarray}
 providing some restrictions to the torsion scalar: since the scale factor $a(t)$ is defined positive, the torsion scalar must satisfy $T<-6\Omega_{\Lambda}^0H_0^2$, and the first modified FLRW equation (\ref{ines18}) yields,
\begin{eqnarray}
- \frac{1}{2}f(T)+\ T\ f_T(T)= \frac{1}{2}T+3\Omega_{\Lambda}^0H_0^2\ ,
\label{lcdm5}
\end{eqnarray}
whose general solution reads
\begin{eqnarray}
f(T)=T+Q\sqrt{-T}-2\Lambda , \label{s1}
\end{eqnarray}
where $Q$ is an integration constant. This kind of model has been proposed in \cite{21desetare}-\cite{24desetare} considering  $f(T)=T-\tilde{\alpha} (-T)^{-n}$ to explain the late-time accelerated expansion without including the dark energy component. In the same way, Bengochea and collaborators \cite{14desetare} investigated the observational information for this model by using the most recent SN Ia+ BAO + CMB data and found that the model can present radiation era, matter era and late-time acceleration phase as the last three phases of cosmological evolution in standard way if $n\in[0.23,0.03]$ and $\Omega_m\in [0.25,0.29].$ Moreover, Setare and collaborator \cite{st5} shown that the model $f(T)=T-\tilde{\alpha}\sqrt{-T}$ presents the stability of de Sitter solution. \par

Now, for consistency, let us use the initial conditions to carry out the exact expression of the constant $Q$. As performed in \cite{condinit,st28}, the algebraic function $f(T)$ has to obey the following initial conditions
\begin{eqnarray}
\left(f\right)_{t=t_i}=T_i,\quad \left(\frac{df}{dt}\right)_{t=t_i}=\left(\frac{dT}{dt}\right)_{t=t_i},\label{arsene1}
\end{eqnarray} 
where $t_i$ denotes the early time at which the initial value of the torsion scalar is called $T_i$. Since one has just a constant in (\ref{s1}), the first initial condition in (\ref{arsene1}) is enough for determining $Q$. By making use of this initial condition, one gets the following value
\begin{eqnarray}
Q= \Omega^{0}_{\Lambda}\frac{H_0^2}{H_i}\sqrt{6},\label{cvalue}
\end{eqnarray}
meaning that $Q$ is directly linked with the cosmological and is always positive. It is straightforward to see that when $Q=0$, the model reduces to the TEGR one. However, when $Q$ does not vanish the model may perfectly reproduce the $\Lambda CDM$ model feature. \par

In what follows, we will assume the ordinary content of the universe as a perfect fluid with an energy density $\rho_m$ and pressure $p_m$, obeying the barotropic equation of state (EoS) $p_m=w_m\rho_m$, $w_m$ being the parameter of EoS. 

\section{ Equilibrium description of thermodynamics for $\Lambda CDM$ model in the framework of $f(T)$ gravity }\label{sec4}

\subsection{Writing down the DE components}\label{sec4.1}

For convenience, let us write Eqs.~(\ref{ines18})-(\ref{ines18'}) in the following forms:
\begin{eqnarray}
 H^2= \frac{\kappa^2}{3f_T} \Big( \bar { \rho} +\rho_m \Big) \label{a1}\,,\\
 \dot{H}=- \frac{\kappa^2}{2f_T} \Big( \bar { \rho} +\rho_m + \bar {P} + P_m \Big) \label{a1'}\,,
\end{eqnarray}
where $\bar{\rho}$ and $\bar{P}$ denote the energy density and pressure of the dark energy, respectively and read
\begin{eqnarray}
   \bar{\rho}&=&\frac{1}{2\kappa^2 }(Tf_T-f) \label{c1}\,,\\ 
\bar{P}&=&\frac{1}{2\kappa^2 } \Big[-(Tf_T -f)+4H f_{TT} \Big] \label{c1'}.
\end{eqnarray}
 Making use of the above equations one gets the following equation 
 \begin{eqnarray}
\dot{ \bar { \rho}} + 3H \Big(\bar{\rho}+\bar{P} \Big)= - \frac{T}{2\kappa^2} f_{TT}  \label{1a}\,,
 \end{eqnarray}
 which leads to the equation of continuity when the rhs  vanishes, i.e, for a linear form of the algebraic function $f(T)$.\\
 We are interested to the thermodynamic properties of  $\Lambda CDM$ model. We will test the validity of both the first and second laws of thermodynamics within this model.

\subsection{Studying the first law according to $\Lambda CDM$ model}\label{sec4.2}

In this section we are interested to check the validity of the first law of thermodynamics within the model under consideration in this work. To do so, let us first consider the following relation 
$ h^{\alpha\beta} \partial_\alpha \bar{r} \partial_\beta \bar{r}=0$, from which the dynamical apparent horizon can be obtained. Here we have $ h_{\alpha\beta}=$diag $(1, -a^2)$.  Since, we are working in the framework of FLRW cosmology, it is easy to show that the apparent horizon takes the following expression 
\begin{eqnarray}
 \bar{r}=\frac{1}{H} \label{a2}\,.
\end{eqnarray}
By deriving this later with respect to the cosmic time, one gets
\begin{eqnarray}
\frac{d\bar{r}}{dt}=- H \dot{H} \bar{r}^3 \label{a2'}\,\,.
\end{eqnarray}
By introducing (\ref{a1'}) in (\ref{a2'}), one obtains:
\begin{eqnarray}
\frac{f_T}{4 \pi G} \frac{d\bar{r}}{dt}= H \bar{r}^3 
\Big(\rho_m+ \bar{\rho}+P_m+ \bar{P}\Big)\label{a3}\,.
\end{eqnarray}
According to the GR the Bekenstein-Hawking horizon
 entropy is governed by the expression \cite{d6} 
\begin{eqnarray}
\bar{S}= \frac{\mathcal{A} f_T}{4G} \label{a3'}\,,
\end{eqnarray}
where $\mathcal{A}$ denotes the area of the apparent horizon . By combining (\ref{a3}) and (\ref{a3'}), one gets
\begin{eqnarray}
\frac{1}{2 \pi \bar{r}} \frac{d\bar{S}}{dt}=
4 \pi H \bar{r}^3 \Big(\rho_m+ \bar{\rho}+P_m+ \bar{P} \Big) +
\frac{\bar{r}}{2G} \frac{df_T}{dt}\,.         \label{a4}  
\end{eqnarray}
Now, making use of the Hawking temperature as
\begin{eqnarray}
 T_H= \frac{|\kappa_{sg}|}{2\pi} \,, \label{a4'}  
\end{eqnarray}
one gets the parameter $\kappa_{sg}$ as follows \cite{a7}
\begin{eqnarray}
 \kappa_{sg}&=&\frac{1}{2\sqrt{-h}} \partial_\alpha 
\Big (\sqrt{-h} h^{\alpha\beta} \partial_\beta \bar{r} \Big) \cr
 &=&- \frac{1}{\bar{r}} \Big (1- \frac{\dot{\bar{r}}}{2H\bar{r}} \Big)\,,
\end{eqnarray}
according to what (\ref{a4'}) becomes
\begin{eqnarray}
T_H=  \frac{1}{2 \pi \bar{r}} \Big(1- \frac{\dot{\bar{r}}}{2H\bar{r}} \Big)\,. \label{a4''} 
\end{eqnarray}
A direct use of (\ref{a4}) and (\ref{a4''}) allows us to obtain 
\begin{eqnarray}
T_H d\bar{S}&=&
4 \pi H \bar{r}^3 \Big(\rho_m+ \bar{\rho}+P_m+ \bar{P} \Big)  dt 
-2\pi \bar{r}^2 \Big(\rho_m+ \bar{\rho}+P_m+ \bar{P} \Big) d\bar{r}+\frac{T_H}{G}\pi \bar{r}^2  df_T    \,.     \label{a5}  
\end{eqnarray}
Instead of the expression $E_{MS}\equiv \bar{r}/(2G)$ for the Misner-Sharp
energy in the GR, 
the suitable expression within modified teleparallel gravity reads
\begin{eqnarray}
\bar{E}_{MS}= \frac{rf_T}{2G} \,. \label{a5'} 
\end{eqnarray}
Making use of (\ref{a2}) and  (\ref{a5'}),   $\bar{E}_{MS}$ becomes
\begin{eqnarray}
\bar{E}_{MS}&=&V \frac{3H^2f_T}{8\pi G} \cr \label{a6} 
&=& V(\rho_m+\bar{\rho})\,,
\end{eqnarray}
where $V$ represents the volume inside the horizon. The first derivative 
of (\ref{a6}) yields
\begin{eqnarray}
 \frac{d\bar{E}_{MS}}{dt}&=&
-4 \pi H \bar{r}^3 \Big(\rho_m+ \bar{\rho}+P_m+ \bar{P} \Big)  
+ 4 \pi \bar{r}^2 \Big(\rho_m+ \bar{\rho} \Big) d\bar{r}+\frac{\bar{r}}{2G}  \frac{df_T}{dt} \,,\label{a6'}  
\end{eqnarray}
from which we deduce
\begin{eqnarray}
T_H d\bar{S}= d\bar{E}_{MS}+ 
2\pi \bar{r}^2
\Big (\rho_f+ \bar{\rho}{_d}-P_f-
  \bar{P}_{d} \Big)
   d\bar{r} 
   + \frac{\bar{r}}{2G}
 \Big(1+ 2\pi \bar{r} T_H \Big) df_T .\label{a7}  
\end{eqnarray}
We can now introduce the work density by 
\begin{eqnarray}
 \bar{W}&=& \frac{-1}{2} \Big( \mathcal{T}_m^{\;\alpha\beta}h_{\alpha\beta}+ 
 \bar{\mathcal{T}}^{\;\alpha\beta}h_{\alpha\beta}\Big) \cr
 &=& \frac{1}{2} \Big(\rho_m+ \bar{\rho}-P_m- \bar{P}\Big)\,,
\end{eqnarray}
where $\bar{\mathcal{T}}^{\;\alpha\beta}$ is the energy momentum tensor related to the dark content. Therefore,
(\ref{a7}) takes the following form
\begin{eqnarray}
T_H d\bar{S}= - d\bar{E}_{MS}+ \bar{W}dV
+ \frac{\bar{r}}{2G} \Big(1+ 2\pi \bar{r} T_H \Big)  df_T,         \label{a7'}  
\end{eqnarray}
which can be rewritten as
\begin{eqnarray}
T_H d\bar{S} +T_H d_i\bar{S}= - d\bar{E}_{MS}+ \bar{W}dV,        \label{a7''}  
\end{eqnarray}
where, after identification, one gets
\begin{eqnarray}
 T_Hd_i\bar{S}&=& - \frac{\bar{r}}{2G} \Big(1+ 2\pi \bar{r} T_H \Big)  df_T \cr
 &=& -T_H\Big(\frac{\bar{E}_{MS}}{T_H}+\bar{S} \Big) \frac{df_T}{f_T} \cr
 &=&\frac{6\pi T_H}{G} \frac{(8HT+\dot{T})}{T(4HT+\dot{T})}  df_T.  \label{a81}                                   
\end{eqnarray}
Observe that the additional term (\ref{a81}) vanishes in
teleparallel theory (as should be the case in GR). However, in a general $f(T)$ theory this term never vanishes and is generally 
interpreted as a production of entropy \cite{a9}. Concerning
our work, we could check the behaviour of this term according to the model
under consideration. To do so, let us make use of the  model (\ref{s1}) and 
check what happens about the addition entropy. \par 
From (\ref{s1}), one can deduce the expression of $f_{TT}$ given by
\begin{eqnarray}
 f_{TT}= \frac{Q}{4T\sqrt{-T}}.
\end{eqnarray}
From equations (\ref{a81}), one gets
\begin{eqnarray}
 \frac{d_i\bar{S}}{dt}&=& \frac{3\pi }{2GT^2} \frac{(8HT+\dot{T})\dot{T}}{(4HT+\dot{T})} 
\frac{Q}{\sqrt{-T}} \nonumber\\
&=& -\frac{Q\pi}{2G\sqrt{6}}\frac{\dot{H}\left(4H^2+\dot{H}\right)}{H^4\left(2H^2+\dot{H}\right)}\,,\label{a81'}                                   
\end{eqnarray}
where we made use of the relation (\ref{m1}) in the second equality of the above equation. Remark that when $Q=0$ the additive term vanishes and the first law according to TEGR is recovered \cite{a7},\cite{a9}. Now, in our case, the constant must be view different from zero in order to preserve the generality of the model. Observe that $4H^2+\dot{H}=\ddot{a}/a+3H^2>0$  and $2H^2+\dot{H}=\ddot{a}/a+H^2>0$, because of the acceleration of the universe. Moreover, the first derivative of the Hubble parameter is positive in an accelerating expanded phantom-like universe and negative in an accelerated expanded quintessence-like one. Therefore, depending on the type of the universe, a phantom or quintessence one, the sign of $d_i\bar{S}/dt$ depends essentially on the sign of the constant $Q$.  Since $Q>0$, in a phantom-like universe, $d_i\bar{S}/dt<0$ ( decreasing entropy) while for a quintessence-like universe one has $d_i\bar{S}/dt>0$ (increasing entropy). Due to the fact that $d_i\bar{S}/dt<0$ in phantom-like universe, the entropy goes to zero as the time evolves. This means that in phantom phase, as the time evolves, the first law of thermodynamics in $f(T)$ gravity reduces to the one in teleparallel gravity.\par
An important feature to be pointed out in this discussion is that, in the case where one has a decreasing additive entropy ($d_i\bar{S}/dt<0$), as the time evolves, the additive entropy tends to zero, where the TEGR situation will be recovered. This allows us to conclude that the production of entropy cannot always be view as permanent phenomenon:  This corresponds to  a phantom-like universe where 
$d_i\bar{S}/dt<0$, as we previously shown.
\subsection{ Studying the second law according to $\Lambda CDM$ model}\label{sec4.3}

In this rubric, we consider the Gibbs equation given by
\begin{eqnarray}
T_H d\bar{S_t}=d\left[\left(\bar{\rho}+\rho_m\right)V \right] + \left(\bar{\rho}+\rho_m+\bar{P}+P_m\right)dV\,.  \label{a8} 
\end{eqnarray}
 Here, $T_H$ and $S_{t}$ denote the temperature and the entropy related to the total energy inside the horizon, respectively. Moreover, in this spirit of equilibrium description, we assume the same temperature between the apparent horizon  and inside it \cite{bamba',43bamba'}. 

The second law of thermodynamics imposes the following condition \cite{a12}
\begin{eqnarray}
 \frac{d\bar{S}}{dt}+\frac{d(d_i\bar{S})}{dt}+\frac{d\bar{S}_{in}}{dt}\geq 0 \,.
 \label{a8'}
\end{eqnarray}
By combining Eqs.~(\ref{a1}),(\ref{a7''}),(\ref{a8}), the rhs of (\ref{a8'})
can be rewritten as 
\begin{eqnarray}
 \frac{d\bar{S}}{dt}+\frac{d(d_i\bar{S})}{dt}+\frac{d\bar{S}_{in}}{dt}= 
 \frac{-3}{4G} \frac{\dot{T}^2f_T}{T^3}\,.  
\end{eqnarray}
 
It is important to note that in this work the physical temperature is  used as the temperature of the apparent horizon, i.e., the Hawking temperature (\ref{a4'}). We also consider in the discussion about the thermodynamics in this modified gravity that the temperature of the universe inside the horizon is equal to that at the apparent horizon, as assumed in \cite{bamba'}. This means that the temperature inside the apparent horizon is assumed to be the same as the one of matter species including that of the CMB photons. We emphasize that in the cosmological setup, the temperature of matter species is about $2.73$K, which can be determined in a standard way \cite{bamba'}.

According to the inequality (\ref{a8'}) one gets the following relation: 
\begin{eqnarray}
  f_T \geq 0\,,\quad\Longrightarrow Q\leq 2H\sqrt{6}.
\end{eqnarray}
 Since it is previously shown that $Q>0$, one gets  $0<Q\leq 2H\sqrt{6}$. Regarding the present
stage of the universe and according to the observational
data, one has $H(t_0)=H_0=2.1\times 0.7\times 10^{-42} GeV$
\cite{a10,a11}, such that we denote the  bound of the constant by $Q_{max}=2H_0\sqrt{6} = 2.94\times 10^{-42} GeV$ for
satisfying the second law of thermodynamics. We consider a scale factor of the type power-law of the cosmic time, i.e, $a=a_0t^{\frac{2}{3(1+\omega_{eff})}}$, where $\omega_{eff}=P_{eff}/\rho_{eff}$ is the effective parameter of EoS and $P_{eff}$ and $\rho_{eff}$ are the effective pressure and energy density, respectively\footnote{ The effective pressure and energy density are easily calculated by $P_{eff}=P_m+\bar{P}$  and $\rho_{eff}=\rho_m+\bar{\rho}$ from the equations  
(\ref{a1})-(\ref{c1'}) assuming an effective running gravitational constant $G_{eff}=G/f_T$, $G$ being the usual gravitational constant.}. In quintessence phase, namely, $-1<\omega_{eff}<-1/3$,  and the exponent $2/(3(1+\omega_{eff}))$ is positive. Since the universe is expanding, we expect a growing scale factor and the cosmic time has to be such that $0^+<t<+\infty$. The Hubble parameter behaves as $H\propto 1/t$, showing that, in the quintessence phase, as the time evolves, the Hubble parameter decreases. Therefore, the current Hubble parameter is less that the initial one, that is, $H_0<H_i$. Hence, writing the constant $Q=\Omega^{0}_{\Lambda}\frac{H_0}{H_i}H_0\sqrt{6}$, the task is just to compare $\Omega^{0}_{\Lambda}\frac{H_0}{H_i}$ with $2$. Due to the fact that $\Omega^{0}_{\Lambda}<1$ and $H_0<1$, it appears clearly that $\Omega^{0}_{\Lambda}\frac{H_0}{H_i}<2$ and then, one always gets $Q<2H_0\sqrt{6}$. Hence, one can conclude that the second law is always satisfied in the quintessence phase.\par
One the other hand, when the phantom phase is considered, one has $2/(3(1+\omega_{eff}))<0$ and in order to realize the expansion, it is necessary to have $-\infty< t<0^-$. In this case, the Hubble parameter 
 grows as the time evolves, meaning that $H_i<H_0$.
Here, it is obvious that the initial time is toward $-\infty$. The Hubble parameter in this case behaves like $H\propto -1/t$ and at early time, $t\rightarrow -\infty$, it goes toward zero, such that the quotient $\frac{H_0}{H_i}$ is very high. Moreover, assuming that the proportion of ordinary matter should be neglected with respect to that of the dark energy (characterized by the cosmological constant and it related terms), one can approximate the parameter $\Omega^{0}_{\Lambda}$ to $1$ and it will appear clearly that $Q>2H_0\sqrt{6}$. Therefore, it can be concluded that the second law of thermodynamics is not satisfied in the phantom phase.

\section{Stability of cosmological solutions}\label{sec5}

This section is devoted to the study of the stability of the reconstructed $\Lambda CDM$ model. To do so, we will insert the homogeneous and isotropic perturbations around the model. In a first time we will work with a general $f(T)$ algebraic function and later, take into account the $\Lambda CDM$. In general, for this kind of work, it is useful to assume the Hubble parameter $H_0(t)=h(t)$, satisfying the modified FLRW cosmological equations (\ref{ines18} - \ref{ines18'}). By solving the equation of continuity (\ref{b1}), one gets the following solution
\begin{eqnarray}
\rho_{mh}(t)=\rho_{o} \; e^{-3\;(1+w_{m})\;\int \;h(t)\; dt}.\label{eq22}
\end{eqnarray}
We now consider small deviations form the Hubble parameter and energy density  in terms of perturbations and write them as follows
\begin{eqnarray}
 H(t)= h(t)\;\Big[1+\delta(t)\Big], \quad \rho_{m}(t)=\rho_{mh}(t)\;\Big[1+\delta_{m}(t)\Big].\label{eq23}
\end{eqnarray}
Here the perturbation functions $\delta(t)$ and $\delta_{m}(t)$ denote the perturbation about the geometry and the matter, respectively. For our purpose  in this work, we will deal with the linear perturbation. Therefore, the algebraic function $f(T)$ can be expanded about the value of the torsion scalar in the background, namely $\tau = -6\;H^{2}(t). $, yielding 
\begin{eqnarray}
 f(T)= f(\tau) + \;f_{T}(\tau)(T-\tau)+\frac{1}{2}\;
 f_{TT}(\tau)(T-\tau)^2+ \mathcal{O}^{3} \label{eq24},
 \end{eqnarray}
$\mathcal{O}^{3}$ characterize the terms of higher power of $T$ which have to be neglected. 
By injecting  (\ref{eq24}) in  (\ref{ines18}) and making  use of the expansion (\ref{eq24}), one gets the following equation  
\begin{eqnarray}
 -6 \;h^{2}(t) \;\Big[ \;f_{T}(\tau)+12\;h^{2}(t)\;f_{TT}(\tau) \Big]\;\delta(t)= 
 \kappa^{2} \;\rho_{mh}(t)\; \delta_{m}(t) \,. \label{eq25}
\end{eqnarray}
This equation points out the relationship between the matter perturbation, the geometric one and also the perturbed Hubble parameter. We clearly see this equation in not enough for getting the analytical expression of the perturbations functions. We then need to take into account another one. The suitable equation to be used  in this way is the equation of  continuity (\ref{b1}). We first perturb it obtaining 
equation (\ref{b1}),
 \begin{eqnarray}
 \dot{\delta}_{m}(t) + 
 3\; (1+w_{m})
 h(t) \;\delta(t)=0 \label{eq26}.
 \end{eqnarray}
Now, equations (\ref{eq25})-(\ref{eq26}) can be combined to yield a one variable equation from which the analytical expressions of the perturbation function may arise. This process will be performed in the upcoming sections where we will assume both de Sitter  and power-law cosmological solutions.

\subsection{Stability of de Sitter solutions}\label{sec5.1}

In de Sitter solutions, the Hubble parameter for the background is a constant and one has
\begin{eqnarray}
h(t) = h_{0} \Longrightarrow a(t)= a_{0} \;e^{h_{0}\;t},  \label{eq217}
\end{eqnarray}
where $h_{0}$ is a constant. Therefore, (\ref{eq22}) becomes
\begin{eqnarray}
\rho_{mh}(t)\;=\rho_{0}\;  e^{-3\;(1+w_{m})\; h_{0}\;t}. \label{eq28}
\end{eqnarray}
On the other hand, by combining Eqs. (\ref{eq25})-(\ref{eq26}), one 
gets the following differential equation
\begin{eqnarray}
 6\; h_0 \;\Big[ f_{T}(\mathcal{T})+12\;h_0^{2}(t)\;f_{TT}(\mathcal{T}) \Big]\; \dot{\delta}_{m}= 
3 \;\kappa^{2}\;(1+w_{m})\; \rho_{mh}(t)\; \delta_{m}(t),  \label{eq29}
\end{eqnarray}
where we made use of (\ref{eq217}). The algebraic function $f(T)$ to be used here is 
that obtained in (\ref{s1}), which is the
 $\Lambda CDM$ model in the context of $f(T)$ gravity. Therefore, by exploring (\ref{s1}) and 
(\ref{eq28}), 
the general solution of (\ref{eq29}) reads
\begin{eqnarray}
 \delta_{m}(t)= K\; \exp \Big[\frac{-\kappa^{2}\; 
 \rho_{0}}{(6\;h_{0}-Q\; \sqrt{6})\;h_{0}} \; e^{-3\;h_{0}(1+w_{m})\; t}\Big], \label{eq30}
\end{eqnarray}
where $K$ is an integration constant. From Eqs.  (\ref{eq25}) and (\ref{eq26}), 
one gets the time evolution of $\delta$ as
\begin{eqnarray}
 \delta(t)= \frac{K\; \kappa^{2}\; \rho_{0}}{(-6\;h^{2}_{0} + Q\; h_{0}\; \sqrt{6})} \; \exp \Big[-3h_{0}(1+w_{m})\;t 
 -
 \frac{\kappa^{2} \;\rho_{0}}{(6\;h_{0}-Q \;\sqrt{6})\;h_{0}} \;  e^{-3\;h_{0}\;(1+w_{m})\;t} \Big]. \label{eq31}
\end{eqnarray}

The functions  $\delta_{m}(t)$ and $\delta(t)$ which allow to study the stability of the model within de Sitter solutions are defined under the condition 
$Q\neq h_0\sqrt{6}$.  Since we are leading with an expanding universe, the constant  $ h_{0}$ is positive.  The parameter of equation of state $w_m$ being related to the ordinary content of the universe, one has $ (1+w_{m})>0 $. Therefore, the function   $\delta_{m} \longrightarrow K $ when $t \longrightarrow \infty$.  $K$ being an integration constant (so an arbitrary constant), we choose it such that 
 $0<K<1$. Then, $\Lambda CDM$ model is stable in the view of matter content within de Sitter solutions.\par
 Concerning the function (\ref{eq31}), one can see that as $t\longrightarrow +\infty$, $\delta \longrightarrow 0$. Hence, we conclude that the model is stable in the view of the geometry. Resuming,  since  $Q\neq h_0\sqrt{6}$, the $\Lambda CDM$ model is stable within de Sitter solutions. An illustrating graph can presented in Figure $1$ showing the convergence of both $\delta_m$ and $\delta$.   More precisely, we perform the evolution of the functions $\delta_m/\delta_{m0}$ and $\delta/\delta_0$ where $\delta_{m0}$ and $\delta_0$  are the current values of the perturbation functions, while $\delta_m$ and $\delta$ are the perturbation function at any time $t>t_0$. We see  from the Fig. $1$  that at $t=t_0$, one has $\delta_m(t_0)/\delta_{m0}=\delta(t_0)/\delta_0=1$, and as the time evolves the curves decrease and goes toward finite values. We assume the current time $t_0$ as the initial time and the curves are plotted in terms of $t-t_0$, where the values of the input parameters are indicated in the legend of each them. The important feature to be carried out here is the impact of the input parameters on the convergence of the curved. It is clear that the curves decrease but one sees that the one linked with the geometrical part converges more rapidly than that characterizing the matter content.  Moreover, one notes that the geometrical  perturbation function fades away for large time, $\delta(t\rightarrow +\infty)\rightarrow 0$, while the matter perturbation function goes asymptotically to a fix value different from zero. This means that, according to the input parameters, the background is more stable, but the matter fluctuations make the matter part less stable.
 
\subsection{Stability of power-law solutions}\label{sec5.2}

In this rubric we are interested to the cosmological solutions of the type
\begin{eqnarray}
 a(t) \propto t^{\alpha} \Longrightarrow h(t)= \frac{\alpha}{t}. \label{eq32}
\end{eqnarray}
By using this later, (\ref{eq25}) becomes
\begin{eqnarray}
\rho_{mh}(t)=\rho_{0}\;  e^{-3\; \alpha\;(1+w_{m})\; \ln\; t}.\label{eq33}
\end{eqnarray}
On the other hand, by combining Eqs. (\ref{eq25})-(\ref{eq26}), one 
gets the following differential equation
\begin{eqnarray}
 6\; h(t)\Big[ f_{T}(\tau)+12h^{2}(t)f_{TT}(\tau) \Big] \dot{\delta}_{m}
 = 
3 \kappa^{2}\;(1+w_{m})\; \rho_{mh}(t) \;\delta_{m}(t).  \label{eq34''}
\end{eqnarray}
The algebraic $f(T)$ to be used here is 
that obtained in (\ref{s1}), which is the
 $\Lambda CDM$ models in the context of $f(T)$ gravity. By making use of (\ref{s1}) and 
(\ref{eq33}), 
the general solution of (\ref{eq34''}) reads
\begin{eqnarray}
 \delta_{m}(t)= K_{1}\; \exp \Big[\frac{B}{\tilde{a}\;(1+m)}\; t^{(1+m)} \; _2F_{1} \Big(1,(1+m),(2+m);\frac{-b\,t}{\tilde{a}} \Big) \Big] \label{eq35}\,,
\end{eqnarray}
with
$$ B= -3\;\kappa^{2}\;\rho_{0}\;(1+w_{m}), \;\;\; m=1-3\;\alpha \;(1+w_{m}),\;
b=Q\;\sqrt{6} \quad and \quad \tilde{a}=-6\;\alpha \,,  $$
where $K_{1}$ is an integration constant. From Eq.  (\ref{eq25}) or (\ref{eq26}), 
one gets the time evolution of $\delta$ as
\begin{eqnarray}
 \delta(t)&=&- \frac{\kappa^{2}\;\rho_{0} \;K_{1}}{\Big(-6\;h^{2}(t)+Q\;h(t)\;\sqrt{6} \Big) } 
 \exp \Big \{-3\;\alpha \;(1+w_{m})\;\ln \;t + \cr
 && \exp \Big [\frac{B\; t^{(1+m)}}{\tilde{a}\;(1+m)}  \;
 _2F_{1} \Big(1,(1+m),(2+m),\frac{-b\;t}{\tilde{a}} \Big) \Big]  \Big \} \,,\label{eq36}
\end{eqnarray}
where the hypergeometric function   $ _2F_{1}$ is defined by  \\
$$ _2F_{1}(\lambda_{1},\lambda_{2},\lambda_{3},z)= 
\sum^{\infty}_{n=0} \frac{(\lambda_{1})_{n}(\lambda_{2})_{n}}{(\lambda_{3})_{n}} \frac{z^{n}}{n!}\,,$$ 
with 
$$(\lambda_{1})_{n}=\lambda_{1}(\lambda_{1}+1)(\lambda_{1}+2)...(\lambda_{1}+n-1), \quad (\lambda_i)_{0}=1   \,.$$
We can rewrite (\ref{eq35}) and (\ref{eq36}) by setting $x=(-b\;t)/a$, obtaining 
\begin{eqnarray}
 \delta_{m}&=& K_{1}\; \exp \Big[\frac{B\; \tilde{a}^{m}\; b^{(-1-m)}\; (-)^{1+m}}{(1+m)}\; x^{(1+m)} \; _2F_{1} \Big(1,(1+m),(2+m),x \Big) \Big] \label{eq351}\,,\\
 \delta &=&- \frac{\kappa^{2}\;\rho_{o}\; K_{1}}{\Big(-6\;h^{2}(x)\;+Q\;\;h(x)\sqrt{6} \Big) } 
 \exp \Big \{-3\;\alpha\; (1+w_{m})\;\ln \;(\frac{-\tilde{a}\;x}{b}) + \cr
 && \exp \Big[\frac{B\; \tilde{a}^{m}\; b^{(-1-m)}\; (-)^{1+m}}{(1+m)}\; x^{(1+m)} \; _2F_{1} \Big(1,(1+m),(2+m),x \Big) \Big]  \Big \}\,, \label{eq361}
\end{eqnarray}
with  $m\neq -1$. An analysis of the asymptotic behaviour of  (\ref{eq351}) and (\ref{eq361}) for $ x\rightarrow\pm \infty$ allows to rewrite them in the following forms:
\begin{eqnarray}
 \delta_{m}&=& K_{1} \;\exp \Big[\frac{B\; \tilde{a}^{m} \;b^{(-1-m)} \;(-)^{2+m}}{(1+m)} \;x^{m} \Big] \label{eq351}\,,\\
 \delta&=&- \frac{\kappa^{2}\;\rho_{0}\; K_{1}}{\Big(-6\;h^{2}(x)+Q\;h(x)\;\sqrt{6} \Big) } 
 \exp \Bigg \{-3\;\alpha\; (1+w_{m})\;\ln \;\left(\frac{-\tilde{a}\;x}{b}\right)  + \cr
 && \exp \Big[\frac{B\; \tilde{a}^{m}\; b^{(-1-m)} \;(-1)^{2+m}}{(1+m)} \;x^{m}\Big] \Bigg \}\,. \label{eq361}
\end{eqnarray}
We start our analysis by the function $\delta_{m}(t)$ which is defined  for 
$m \neq -1$, $Q \neq 0$ and $ t \neq 6 \alpha/(Q\sqrt{6})$. The whole analysis will be done distinguishing the following two sub-cases:\\

\begin{enumerate}
 
\item For  $m<0$  it is easy to see that  $\delta_{m}(t) \longrightarrow  K_{1}
 $ when $t \longrightarrow \infty$. Therefore, we choose this integration function such that  $0<K_{1}<1$. Here the matter stability is always ensured;
 
\item For $m>0$, and an even  $m$, the function  $\delta_{m}(t)$ diverges as 
 $t$ est enough high leading to an instability. However, for an odd $ m $ the function  $\delta_{m}(t) \longrightarrow 0$. In this case, there is stability;


\end{enumerate}

Concerning the function $\delta(t)$, the suitable conditions of it definition read
$m \neq -1$, $Q \neq 0$ and $ t \neq \frac{6 \alpha}{Q\sqrt{6}}$.

\begin{enumerate}

 
\item  For  $ m>0$ and an odd $ m $,  $\delta(t)$ diverges (instability), whereas for an even $m$ it tends to zero (stability);

 \item For  $m<0$  the perturbation function  $\delta(t)$ goes to zero (stability).

\end{enumerate}

Here, it appears that, in contrast to what happens in de Sitter case, where the solutions are always stables independently of the value of the constant $Q$, we see that in the case of power-law solutions, depending on the value of $Q$, instability may appear. This fact is completely predictable and is linked with the behaviour of the ordinary energy density. The physical reason to this is the convergence presented by the quotient $\frac{\rho_{hm}}{\rho_0}$,  the energy density at any time $t>t_0$ over the one at the current time $t_0$. This reflects the fact that as the time evolves, the ordinary matter content reduces, whereas the dark content grows. Let us show prove it in a more clearly way. First, since the parameter of the EoS of the ordinary matter content $\omega$ is always positive, it is easy to observe from (\ref{eq28}) that, as the cosmic time evolves, the ordinary energy density goes to zero. From Eq.~(\ref{eq28}), the cosmic time may be extracted such that
\begin{eqnarray}
\ln{\left(\frac{\rho_{hm}}{\rho_0}\right)}=-3h_0\left(1+\omega_m\right)t.\label{proof1}
\end{eqnarray}
Once again one can observe in a quantitative way that as the time evolves, the quantity $-3h_0\left(1+\omega\right)$ is more negative, meaning that the quotient $\rho_{hm}/\rho_0$ is much less that $1$ for large time, i.e, $\rho_{hm}<<\rho_0$, which is a physical aspect. Now, how does this reflect on the stability of the de Sitter solutions and not on the power-law ones? This is clear when we substitute the rhs of (\ref{proof1}) in the perturbations functions $\delta_{m}(t)$ and $\delta(t)$ in the de Sitter case, getting
\begin{eqnarray}
\delta_m(t)&=&K\exp{\left[  \frac{-\kappa^2\rho_0}{\left(6h_0-Q\sqrt{6}\right)h_0}\frac{\rho_{hm}(t)}{\rho_0} \right]}\\
\delta(t)&=&\frac{K\kappa^2\rho_0}{h_0\left(-6h_0+Q\sqrt{6}\right)}\exp{\left[\ln{\left(\frac{\rho_{hm}}{\rho_0}\right)}-\frac{\kappa^2\rho_0}{\left(6h_0-Q\sqrt{6}\right)h_0}\frac{\rho_{hm}(t)}{\rho_0} \right]}.
\end{eqnarray}
Remark that from the value of $Q$ in (\ref{cvalue}), one always has $\left(6h_0-Q\sqrt{6}\right)\neq 0$, meaning that the above perturbation functions are always defined. Moreover, as the cosmic time evolves, i.e, $\rho_{hm}/\rho_0<<1$, one falls in the situation where $\delta_m(t)<<1$ and $\delta(t)<<1$ by suitably adjusting the papers. This stability of the perturbation functions comes from the fact that these functions exclusively depend one the  cosmological parameters related to ordinary matter.\par
One the other hand, using Eqs.~(\ref{eq33}), (\ref{eq35})-(\ref{eq36}), the perturbation functions for power-law solutions read
\begin{eqnarray}
\delta_{m}(t)&=& K_{1}\; \exp \Big[\frac{B}{\tilde{a}\;(1+m)}\; \left(\frac{\rho_{mh}}{\rho_0}\right)^{-\frac{1+m}{3\alpha(1+\omega)}} \; _2F_{1} \Big(1,(1+m),(2+m);\frac{-b}{\tilde{a}} \left(\frac{\rho_{mh}}{\rho_0}\right)^{-\frac{1}{3\alpha(1+\omega)}}\Big) \Big]\\
 \delta(t)&=&- \frac{\kappa^{2}\;\rho_{0} \;K_{1}}{\Big(-6\;h^{2}(t)+Q\;h(t)\;\sqrt{6} \Big) } 
 \exp \Big \{\ln{\left(\frac{\rho_{mh}}{\rho_0}\right)} + \cr
 && \exp \Big [\frac{B}{\tilde{a}\;(1+m)} \left(\frac{\rho_{mh}}{\rho_0}\right)^{-\frac{1+m}{3\alpha(1+\omega)}} \;
 _2F_{1} \Big(1,(1+m),(2+m),\frac{-b}{\tilde{a}} \left(\frac{\rho_{mh}}{\rho_0}\right)^{-\frac{1}{3\alpha(1+\omega)}}\Big) \Big]  \Big \}
\end{eqnarray}
In this present case, one can observe that besides the ordinary content parameters, these functions also depend on the dark content parameters. More precisely the effect of the dark content is incorporated in the parameter $\alpha=2/(3(1+\omega_{eff}))$, where $\omega_{eff}$ is the parameter of effective EoS, as previously defined in the subsection $4.3$. One sees that the convergence or not of the perturbations functions depends on the sign of $\alpha$. Therefore, one can distinguish two interesting stages of the universe, according to which the stability may appear: the phantom and quintessence phases. So, if the universe lives the phantom phase, $\omega_{eff}<-1$, the parameter $\alpha<0$, $\tilde{a}>1$ and $(1+m)>0$, such that, as the time evolve, i.e, $\rho_{hm}/\rho_0<<1$, the argument of the hypergeometric functions goes to zero and the perturbation functions converge. However, in the quintessence phase, $\alpha>0$, the perturbation  functions may diverge and then the stability is not always guaranteed. We see clearly from all this that, the (in)stability of power-law solutions does not depend only of the ordinary content of the universe, but also on the dark ones. One can then conclude that the stability of the power-law solutions  within the model being studied here depends on the stage in which the universe is living, phantom or quintessence. In the phantom phase, power-law solution present stability, while for the quintessence universe, the solutions are unstable.\par
An interesting question to be answered is, do power-law solutions are stable in the limit $\alpha\rightarrow \infty$? This question can be answered by explicitly considering the expression of the parameter $\alpha$, that is, $2/(3(1+\omega_{eff}))$. Observe that for both de Sitter and power-law solutions, the expressions of the perturbation functions are essentially base on that of the scale factors. Therefore, if an application can transform one of them to the other, a direct consequence can be deduced about their respective perturbations. More precisely, it is well known that for de Sitter solutions, the parameter of EoS is $\omega_{eff}=-1$, $P_{eff}=-\rho_{eff}$. On the other hand, from the scale factor of the power-law solutions (\ref{eq32}), $\alpha$ goes to $\infty$ if and only if $\omega_{eff}=-1$, corresponding to de Sitter solutions. The same question should be answered by extracting the time $t$ from (\ref{eq32}) and injecting it in (\ref{eq35})-(\ref{eq36}) and later, taking the limit $\alpha\rightarrow \infty$. It is straightforward to observe that the perturbation functions converge.

\par

Now, in order to point out the suitable condition for getting
a whole stability we need to take into account  both
the above enumerative conditions.  
After an appropriate analysis,
we see that for any $m<0$, being odd or even, the stability of $\Lambda CDM$ model 
is always ensured.  This is illustrated in Fig. $2$, where both $\delta_m$ and $\delta$ converge as the time evolves. As performed in the de Sitter case, we  also present the evolution of the quotients $\delta_m/\delta_{m0}$ and $\delta/\delta_0$, where, once again, we see that the geometrical perturbation function is more stable than the one of the matter content. Also, it is numerically obvious that the geometrical perturbation fade away asymptotically, while the matter perturbation continues existing but with a fix value for large time.  For any other condition different from the above one, the model turn out to be unstable.\par
 
According to the two cosmological solutions (de Sitter and power-law solutions), 
the sufficient condition for which the $\Lambda CDM$ model in the context of $f(T)$ gravity
is stable is $m<0$ ($\alpha >1/3(1+w_m)$). \par
 On the other hand, because of the fact that $Q<Q_{max}$, it is easy to observe that the first and second
laws are simultaneously satisfied for $\alpha >1/3(1+w_m)$.

 \begin{figure}[hbtp]
 \includegraphics[scale=1]{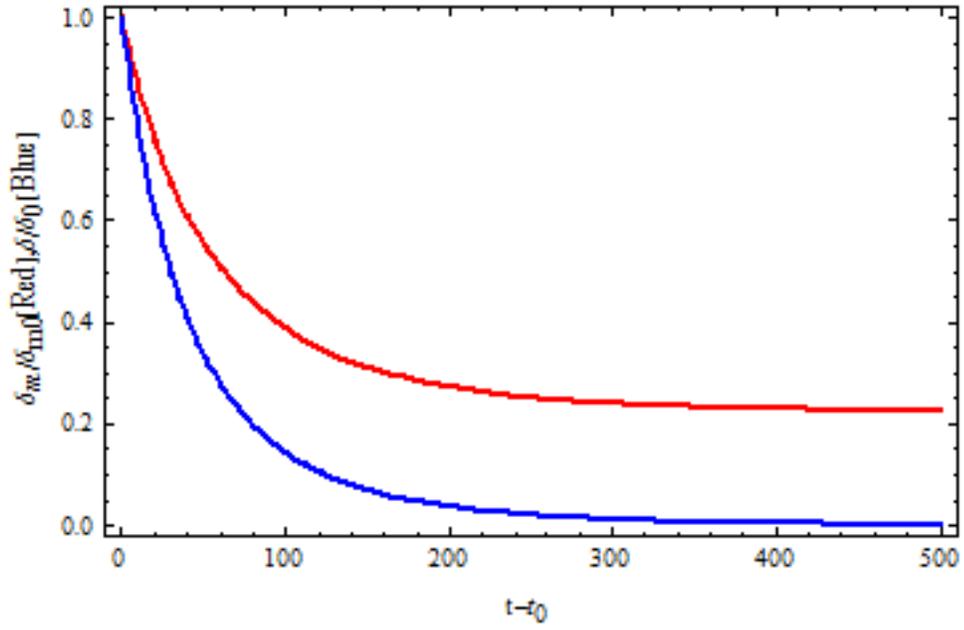}
 \caption{The graphs illustrating the evolution of the perturbation functions $\delta_m $ and $\delta$ as the time evolves in the case of de Sitter solutions, showing their convergence (stability of the model). The function are plotted for 
$ K=0.5$, \,\, $\rho_0= 0.1\times 10^{-121}$,
 \, $Q=Q_{max}/3$, \, $w_m=0$ ,\, $h_0=2.1\times 0.7\times 10^{-42}$.
 }
 \end{figure}


\begin{figure}[hbtp]
\includegraphics[scale=1]{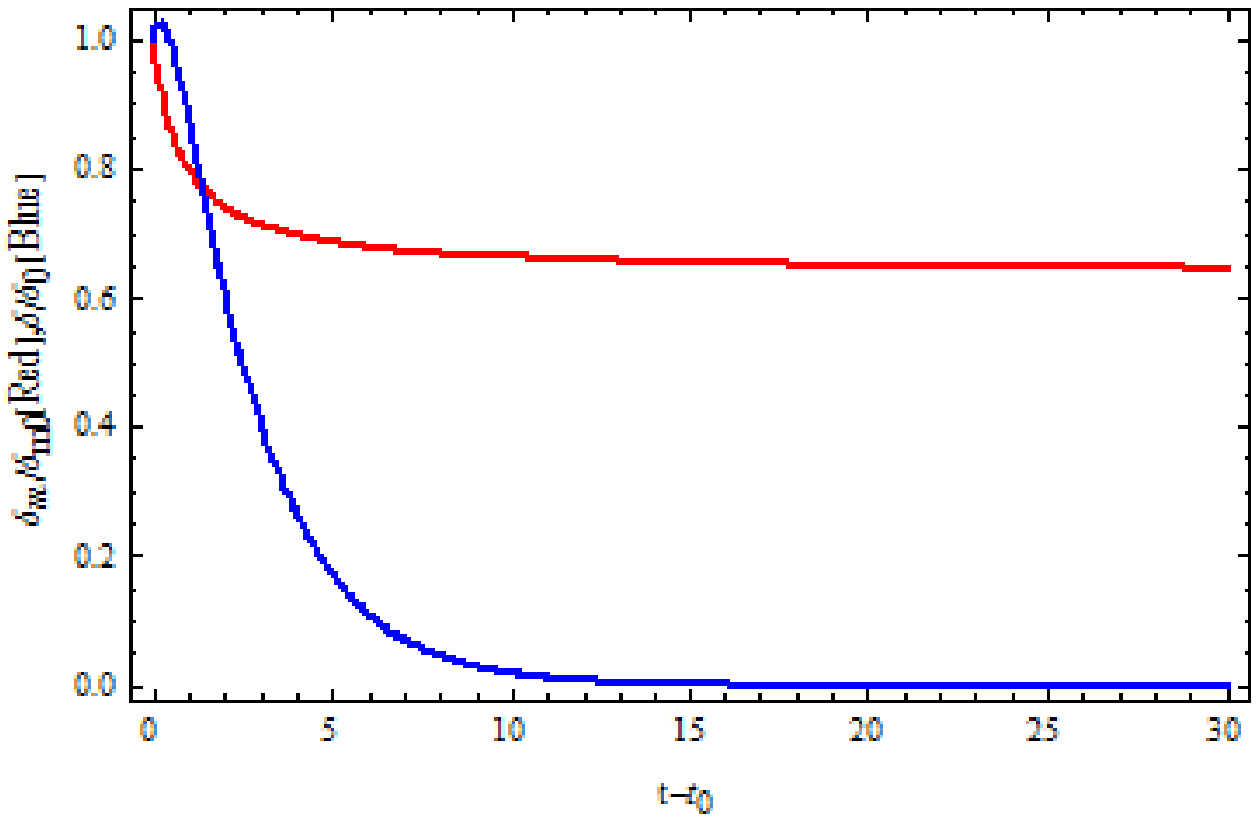}
\caption{The graph illustrating the convergence  of the perturbation functions $\delta_m $ and $\delta$ as the time evolves in the case of power-law solutions. The function are plotted for 
$ K=0.5$, \,\, $\rho_0= 0.1\times 10^{-121}$,
 \, $Q=Q_{max}/3$, \, $w_m=0$, \, 
 \, $m=5$, $\alpha=-2$,\, $h_0=2.1\times 0.7\times 10^{-42}$. }
\end{figure}

 \section{Conclusion}\label{sec6}
This paper is devoted to some features of the $\Lambda CDM$ model if the framework of $f(T)$ modified theory of gravity.  We first perform the reconstruction of $f(T)$ model according to mathematical aspect of the so-called $\Lambda CDM$. The model present an integration constant $Q$ which played and important role in the study of the thermodynamics description and the stability of the model.   In the view of thermodynamics and in order to carry out the effect of the modified algebraic function $f(T)$, we focused our attention on the equilibrium description of the thermodynamics, where we assumed that the temperature of the universe inside the horizon is equal to that of the apparent horizon.\par
We shown that the first law of thermodynamics with an increasing additive entropy is realized only in quintessence-like universe. The first continue occurring  in phantom-like universe but as the time evolves, this additive entropy fades away and the teleparallel aspect is recovered. The second law is satisfied only in quintessence phase.\par
Moreover, in order to check the viability of this model, we study its stability taking into account the de Sitter and power-law cosmological solutions. The results show that within the de Sitter solutions, the stability is always realized, whereas in the power-law one, the stability is obtained depending of the  type of universe, phatom or quintessence-like. We see that in phantom phase, the stability is always satisfied, whereas, in quintessence-like universe solutions present instability.\par
Furthermore, we mix the thermodynamics aspect and the stability one and point out the constraint on the integration constant for obtaining a viability of the model (at least within these considerations). Therefore, we observe that both thermodynamics laws and stability are realized only for a quintessence-like universe within de Sitter solutions.

\vspace{1cm}

{\bf Acknowledgement}: Authors thank Prof. S. D. Odintsov for useful comments and discussions.  I. G. SALAKO thanks  ICTP/IMSP for partial financial support. M. E. Rodrigues thanks UFPA  for the hospitality during the realization of this work and also CNPq for financial support.


\begin{thebibliography}{17}

\addcontentsline{toc}{chapter}{Bibliographie}
\bibitem{debamba1}
S. Perlmutter et al. [SNCP Collaboration], Astrophys. J. {\bf 517}, 565 (1999); A. G. Riess et al.[SNST Collaboration], Astron. J. {\bf 116}, 1009 (1998).
\bibitem{debamba2}
D. N. Spergel et al. [WMAP Collaboration], Astrophys. J. Suppl. {\bf 148}, 175 (2003); ibid. 170,
377 (2007); E. Komatsu et al. [WMAP Collaboration], ibid. {\bf 180}, 330 (2009).
\bibitem{debamba3}
E. Komatsu et al. [WMAP Collaboration], Astrophys. J. Suppl. {\bf 192}, 18 (2011).
\bibitem{debamba4}
M. Tegmark et al., Phys. Rev. D {\bf 69}, 103501 (2004); U. Seljak et al. [SDSS Collaboration],
Phys. Rev. D {\bf 71}, 103515 (2005).
\bibitem{debamba5}
D. J. Eisenstein et al., Astrophys. J. {\bf 633}, 560 (2005).
\bibitem{debamba6}
B. Jain and A. Taylor, Phys. Rev. Lett. {\bf 91}, 141302 (2003).

\bibitem{a'}
R.Aldrovandi and J.G.Pereira,TELEPARALLEL GRAVITY, \\in
http://www.ift.unesp.br/users/jpereira/tele.pdf.
\bibitem{ma1}{\bf A. De Felice and S. Tsujikawa, Living Rel. Rev. {\bf 13}, 3 (2010) [arXiv:1002.4928
[gr-qc]]; K. Bamba, S. Capozziello, S. Nojiri, S. D. Odintsov.  Astrophys. Space Sci. {\bf 342}, 155 (2012) [arXiv:1205.3421 [gr-qc]]};
S. Nojiri and S. D. Odintsov,  	ECONF C {\bf 0602061}, 06 (2006); Int. J. Geom. Meth. Mod. Phys. {\bf 4}, 115-146 (2007) [arXiv:hep-th/0601213];   Phys. Rept. {\bf 505}, 59-144 (2011) [arXiv:1011.0544].

\bibitem{premierfrt}
T. Harko, F. S. N. Lobo, S. Nojiri and S. D. Odintsov, “f(R, T ) gravity,” Phys. Rev. D {\bf 84} (2011) 024020. [arXiv:1104.2669 [gr-qc]].


\bibitem{ma2}
 M. J. S. Houndjo, Int. J. Mod. Phys. D. {\bf 21}, 1250003 (2012). arXiv: 1107.3887 [astro-ph.CO].
 
\bibitem{ma3} 
 M. J. S. Houndjo and O. F. Piattella��, Int. J. Mod. Phys. D. {\bf 21}, 1250024 (2012). arXiv: 1111.4275 [gr.qc].
 
 \bibitem{ma4}
  D. Momeni, M. Jamil and R. Myrzakulov, Euro. Phys. J. C {\bf 72}, arXiv: 1107.5807[physics.gen-ph].
  
\bibitem{ma5}  
 M. J. S. Houndjo, C. E. M. Batista, J. P. Campos and O. F. Piattella,�� [arXiv:1203.6084 [gr-qc]].
\bibitem{ma6}
F. G. Alvarenga, M. J. S. Houndjo, A. V. Monwanou and Jean. B. Chabi-Orou,�� arXiv: 1205.4678 [gr-qc].







\bibitem{mj1}  S.~'i.~Nojiri and S.~D.~Odintsov,
   ``Modified Gauss-Bonnet theory as gravitational alternative for dark
energy,''
   Phys.\ Lett.\ B {\bf 631}, 1 (2005)  [hep-th/0508049]; 
S. Nojiri, S. D. Odintsov, A. Toporensky, P. Tretyakov, arXiv:0912.2488.
\bibitem{mj2}
K. Bamba, S. D. Odintsov, L. Sebastiani, S. Zerbini, arXiv:0911.4390. 
\bibitem{mj3}
K. Bamba, C.-Q. Geng, S. Nojiri, S. D. Odintsov, arXiv:0909.4397.
\bibitem{mj4}
M.E. Rodrigues, M.J.S. Houndjo, D. Momeni, R. Myrzakulov, arXiv:1212.4488.
\bibitem{mj5}
M. J. S. Houndjo, M. E. Rodrigues, D. Momeni, R. Myrzakulov .
 arXiv:1301.4642 [gr-qc].

\bibitem{st1} J.~Amor\'os, J.~de Haro and S.~D.~Odintsov,
   ``Bouncing Loop Quantum Cosmology from $f(T)$ gravity,''
   Physical Review D 87, {\bf 104037} (2013)
   [arXiv:1305.2344 [gr-qc]]; 

  K.~Bamba, J.~de Haro and S.~D.~Odintsov,
   ``Future Singularities and Teleparallelism in Loop Quantum Cosmology,''
   JCAP {\bf 1302} (2013) 008
   [arXiv:1211.2968 [gr-qc]];

  K.~Bamba, S.~'i.~Nojiri and S.~D.~Odintsov,
   ``Effective $f(T)$ gravity from the higher-dimensional Kaluza-Klein and
Randall-Sundrum theories,''
   arXiv:1304.6191 [gr-qc]; 

G. R. Bengochea, R. Ferraro and , Phys. Rev. D {\bf 79}, 124019 (2009) [arXiv:0812.1205 [astro-ph]].
\bibitem{st2}
 E. V. Linder, Phys.Rev. D {\bf 81}, 127301 (2010) [Erratum-ibid. D 82, 109902 (2010)] [arXiv:1005.3039 [astro-ph.CO]].
\bibitem{st3}
 M. Jamil, D. Momeni and R. Myrzakulov, Eur. Phys. J. C {\bf 72} (2012) 2267 [arXiv:1212.6017 [gr-qc]].
\bibitem{st4} 
  R. Myrzakulov, Entropy {\bf 14} (2012) 1627[arXiv:1212.2155 [gr-qc]].
  
\bibitem{st5}  
   M. R. Setare and N. Mohammadipour, JCAP {\bf 1211} (2012) 030 [arXiv:1211.1375 [gr-qc]].
\bibitem{st6}   
    M.R. Setare, N. Mohammadipour, JCAP {\bf 01} (2013) 015 [arXiv: 1301.4891].
\bibitem{st7}
    
 M. Jamil, D. Momeni, R. Myrzakulov and P. Rudra, J. Phys. Soc. Jap. {\bf 81}  (2012) 114004 [arXiv:1211.0018 [physics.gen-ph]].
 \bibitem{st8}
  M. E. Rodrigues, M. J. S. Houndjo, D. Saez-Gomez and F. Rahaman, Phys. Rev. D {\bf  86} (2012) 104059 [arXiv:1209.4859 [gr-qc]].
 \bibitem{st9} 
 M. Jamil, D. Momeni and R. Myrzakulov, Eur. Phys. J. C {\bf 72} (2012) 2122 [arXiv:1209.1298 [gr-qc]].
 \bibitem{st10}
  R. Myrzakulov, Eur. Phys. J. C {\bf 72} (2012)2203 [arXiv:1207.1039 [gr-qc]].
  \bibitem{st11}
   M. J. S. Houndjo, D. Momeni and R. Myrzakulov, Int. J. Mod. Phys. D {\bf 21} (2012)
1250093 [arXiv:1206.3938 [physics.gen-ph]].
\bibitem{st12}
 M. E. Rodrigues, M. H. Daouda and M. J. S. Houndjo, arXiv:1205.0565
[gr-qc].
\bibitem{st13}
 M. R. Setare and M. J. S. Houndjo, arXiv:1203.1315 [gr-qc].
 \bibitem{st14}
  K. Bamba, M. Jamil, D. Momeni and R. Myrzakulov, arXiv:1202.6114 [physics.gen-ph].
 \bibitem{st15} 
 K. Bamba, R. Myrzakulov, S. 'i. Nojiri and S. D. Odintsov, Phys. Rev. D {\bf 85} (2012)104036 [arXiv:1202.4057 [gr-qc]].
 \bibitem{st16}
  M. Jamil, D. Momeni and R. Myrzakulov, Eur. Phys. J. C {\bf 72} (2012)  2267 [arXiv:1212.6017[gr-qc]].
  \bibitem{st17}
   M. Jamil, D. Momeni and R. Myrzakulov, Gen. Rel. Grav. {\bf 45} (2013) 263 [arXiv:1211.3740 [physics.gen-ph]].
   \bibitem{st18}
M. Jamil, D. Momeni and R. Myrzakulov, Eur. Phys. J. C {\bf 72} (2012) 2122 [arXiv:1209.1298 [gr-qc]].
\bibitem{st20}

 M. Jamil, D. Momeni and R. Myrzakulov, Eur. Phys. J. C {\bf 72} (2012) 2075 [arXiv:1208.0025 [gr-qc]].
 \bibitem{st21}
  M. Jamil, K. Yesmakhanova, D. Momeni and R. Myrzakulov, Central Eur. J. Phys. {\bf 10} (2012) 1065 [arXiv:1207.2735 [gr-qc]].
  \bibitem{st22}
   M. J. S. Houndjo, D. Momeni and R. Myrzakulov, Int. J. Mod. Phys. D {\bf 21} (2012) 1250093 [arXiv:1206.3938 [physics.gen-ph]].
   \bibitem{st23}
 M. Jamil, D. Momeni and R. Myrzakulov, Eur. Phys. J. C {\bf 72} (2012) 1959 [arXiv:1202.4926 [physics.gen-ph]].
 \bibitem{st24}
 
  M. H. Daouda, M. E. Rodrigues and M. J. S. Houndjo, Phys. Lett. B {\bf 715} (2012) 241 [arXiv:1202.1147 [gr-qc]].
 \bibitem{st25} 
 M. Jamil, S. Ali, D. Momeni, R. Myrzakulov and Eur. Phys. J. C {\bf  72}, 1998 (2012) [arXiv:1201.0895 [physics.gen-ph]].
 \bibitem{st26}
 
  M. Jamil, D. Momeni, N. S. Serikbayev, R. Myrzakulov and , Astrophys. Space Sci.{ \bf 339}, 37 (2012) [arXiv:1112.4472 [physics.gen-ph]].
  \bibitem{st27}
  
   M. Jamil, D. Momeni, M. A. Rashid and , Eur. Phys. J. C {\bf  71}, 1711 (2011) [arXiv:1107.1558 [physics.gen-ph]].
   \bibitem{st28}
 M. Hamani Daouda, M. E. Rodrigues and M. J. S. Houndjo, Eur. Phys. J. C {\bf 72} (2012) 1893 [arXiv:1111.6575 [gr-qc]].
 \bibitem{st29}
 
  M. Hamani Daouda, M. E. Rodrigues and M. J. S. Houndjo, Eur. Phys. J. C {\bf 72} (2012) 1890 [arXiv:1109.0528 [physics.gen-ph]].
  \bibitem{st30}
   R. Myrzakulov, Gen. Rel. Grav. {\bf 44} (2012) 3059 [arXiv:1008.4486 [physics.gen-ph]].
   \bibitem{st31}
 K. K. Yerzhanov, S. .R. Myrzakul, I. I. Kulnazarov and R. Myrzakulov, arXiv:1006.3879 [gr-qc].
 \bibitem{st32}
 
  R. Myrzakulov, Eur. Phys. J. C {\bf 71} (2011) 1752 [arXiv:1006.1120 [gr-qc]].
  \bibitem{st33}
  M. E. Rodrigues, M. J. S. Houndjo, D. Momeni, R. Myrzakulov and , arXiv:1302.4372 [physics.gen-ph].
\bibitem{st34}
 J. M. Bardeen, B. Carter, S. W. Hawking , Commun. Math. Phys. {\bf 31} (1973) 161-170.
 \bibitem{st35}
 
 N. Tamanini and C. G. Boehmer, Phys. Rev. D {\bf 86}, 044009 (2012), arXiv:1204.4593 [gr-qc].
 \bibitem{st36}
 
 Baojiu Li, T. P. Sotiriou and J. D. Barrow, Phys. Rev. D {\bf 83}, 064035 (2011); Phys. Rev. D {\bf 83}, 104030 (2011).
 \bibitem{st37}
 M. J. S. Houndjo, D. Momeni, R. Myrzakulov and M. E. Rodrigues,arXiv:1304.1147.

\bibitem{st38}
 C. Deliduman and B. Yapiskan, arXiv:1103.2225v3 [gr-qc].
\bibitem{st39}
 M. Hamani Daouda, M. E. Rodrigues and M. J. S. Houndjo, Eur. Phys. J. C {\bf 71} (2011) 1817 [arXiv:1108.2920 [astro-ph.CO]].
 
 \bibitem{bambaa1}  
J. M. Bardeen, B. Carter and S. W. Hawking, Commun. Math. Phys. {\bf 31}, 161 (1973); J. D. Bekenstein, Phys. Rev. D {\bf 7}, 2333 (1973); S. W. Hawking, Commun. Math. Phys. {\bf 43}, 199 (1975) [Erratum-ibid. {\bf 46}, 206 (1976)]; G. W. Gibbons and S. W. Hawking, Phys. Rev. D {\bf 15}, 2738 (1977). 
\bibitem{bamba'}
K. Bamba and C. Q. Geng, JCAP {\bf 1111}, 008 (2011) [arXiv:1109.1694 [gr-qc]].
\bibitem{bamb1}
T. Jacobson, Phys. Rev. Lett. {\bf 75}, 1260 (1995). E. Elizalde and P. J. Silva, Phys. Rev. D {\bf 78}, 061501 (2008).

\bibitem{bamb2}
 K. Bamba, C. Q. Geng, S. Nojiri
and S. D. Odintsov, Europhys. Lett. 89, 50003 (2010); S. F. Wu, B. Wang, X. H. Ge and G. H. Yang, Phys. Rev. D 81, 044010 (2010); Y. Yokokura, arXiv:1106.3149 [hep-th].
\bibitem{bamb3}
C. Eling, R. Guedens and T. Jacobson, Phys. Rev. Lett. {\bf 96}, 121301 (2006).
\bibitem{lambda}
A. de la Cruz-Dombriz and A. Dobado, Phys. Rev. D {\bf 74}, 087501 (2006), gr-qc/0607118; S. ’i. Nojiri and S. D. Odintsov,
J. Phys. Conf. Ser. {\bf 66} (2007) 012005 [hep-th/0611071]. S. Nojiri, S. D. Odintsov and D. S´aez-G´omez, Phys. Lett. B 681,
74 (2009), arXiv:0908.1269 [hep-th]; P. K. S. Dunsby, E. Elizalde, R. Goswami, S. Odintsov and D. S´aez-G´omez, Phys.
Rev. D {\bf 82}, 023519 (2010), arXiv:1005.2205 [gr-qc].

\bibitem{a1}
L. Baojiu, T. P. Sotiriou, and J.D. Barrow, Phys. Rev. D {\bf 83}, 064035 (2011);\\ Phys.Rev.D {\bf 83}:104030(2011).
\bibitem{a2}
M. Hamani Daouda, Manuel E. Rodrigues, M. J. S. Houndjo, Eur. Phys. J. C. {\bf 71} 1817 (2011),\\
arXiv:1108.2920v4 [astro-ph.CO]; Euro. Phys. J. C {\bf 72} 1890 (2012).
\bibitem{a3}
M. Hamani Daouda, Manuel E. Rodrigues and M. J. S. Houndjo, arXiv:1205.0565v1 [gr-qc].
\bibitem{a4}
N. Tamanini C. G. Boehmer, Phys. Rev. D {\bf 86}, 044009 (2012) [arXiv:1204.4593 [gr-qc]].
\bibitem{a5}
R. Ferraro and F. Fiorini, Phys. Lett. B {\bf 702}, 75 (2011).

\bibitem{manu1}
Rong-Xin Miaoa, Miao Lib, and Yan-Gang Miaoc arxiv.org/pdf/1107.0515v3.pdf
\bibitem{manu2}
K. Karami, A. Abdolmaleki arxiv.org/pdf/1201.2511v2.
\bibitem{d6}
J. M. Bardeen, B. Carter and S. W. Hawking, Commun. Math. Phys. {\bf 31}, 161 (1973); \\
J. D. Bekenstein, Phys. Rev. D {\bf 7}, 2333 (1973); S. W. Hawking, Commun. Math. Phys. {\bf 43},199 (1975) [Erratum-ibid. {\bf 46}, 206 (1976)]; \\G. W. Gibbons and S. W. Hawking, Phys. Rev. D {\bf 15}, 2738 (1977).
\bibitem{a7}
R. G. Cai and S. P. Kim, JHEP {\bf 0502}, 050 (2005).

\bibitem{a9}
M. J. S. Houndjo, F. G. Alvarengaa, Manuel E. Rodrigues, Deborah F. Jardim and R. Myrzakulov  [arXiv:1207.1646V2 [gr-qc]].



\bibitem{a10}
 E.W. Kolb and M. S. Turner, The Early Universe (Addison-Wesley, Redwood City, California, 1990).

\bibitem{a11}
E. Komatsu et al. [WMAP Collaboration], Astrophys. J. Suppl. {\bf 192}, 18 (2011) [arXiv:1001.4538[astro- ph.CO]].
\bibitem{a12}
Kazuharu Bamba and Chao-Qiang Geng, arXiv:1109.1694v2 [gr-qc].


\bibitem{21desetare} B. Li, T. P. Sotiriou and J. D. Barrow, Phys. Rev. D {\bf 83}, 104017 (2011), arXiv: 1103.2786 [astro-ph.CO].

\bibitem{24desetare} P. Wu and H. Yu, Phys. Lett. B {\bf 692}, 176-179 (2010), arXiv: 1007.2348 [astro-ph.CO]; R. J. Yang, arXiv: 1007.3571 [gr-qc].

\bibitem{14desetare} G. R. Bengochea and R. Ferraro, Phys. Rev. D {\bf 79}, 124019 (2009); G. R. Bengochea, Phys. Lett. B, {\bf 695}, 405 (2011).



\bibitem{condinit}  Xing Wu and Zong-Hong Zhu, Phys. Lett. B {\bf 660}, 293-298 (2008).


\bibitem{43bamba'} S. F. Wu, B. Wang, G. H. Yang and P. M. Zhang, Class. Quant. Grav.{\bf 25}, 235018 (2008).

\end{thebibliography}
\end{document}